# Power-efficient soliton microcombs

Óskar B. Helgason, Marcello Girardi, Zhichao Ye, Fuchuan Lei, Jochen Schröder and

Victor Torres-Company*

Department of Microtechnology and Nanoscience, Chalmers University of Technology, SE-41296

Gothenburg, Sweden

* Correspondence to: torresv@chalmers.se

**Laser frequency combs are enabling some of the most exciting scientific endeavours in the 21st century, ranging from the development of optical clocks to the calibration of the astronomical spectrographs used for searching Earth-like exoplanets[1,2]. Today, dissipative Kerr solitons generated in microresonators[3] offer the prospect of attaining frequency combs in miniaturized systems by capitalizing on advances in photonic integration[4–6]. Most of the applications based on soliton microcombs rely on tuning a continuous-wave laser into a longitudinal mode of a microresonator whose dimensions are engineered to display anomalous dispersion at the pump laser frequency[7]. In this configuration, however, nonlinear physics precludes from attaining dissipative Kerr solitons with high power conversion efficiency[8,9], with typical comb powers amounting to ~1% of the available laser power. Here, we demonstrate that this fundamental limitation can be overcome by inducing a controllable frequency shift to a selected cavity resonance. Experimentally, we realize this shift using two linearly coupled anomalous-dispersion microresonators (a photonic molecule[10]), resulting in a coherent dissipative Kerr soliton with a conversion efficiency exceeding 50% and excellent line spacing stability. We describe the physical soliton dynamics in this configuration, and discover the system displays unusual characteristics, such as the possibility to backwards initiate solitons and stable operation with a blue detuned pump laser. By optimizing the microcomb power available on chip, these results facilitate the practical implementation of a scalable integrated photonic architecture for energy-efficient applications.**

Driven by a continuous wave (CW) laser, dissipative Kerr solitons (DKSs) are maintained through a balance of optical losses with parametric gain, and dispersion with Kerr nonlinearity[3]. DKSs result in a train of optical pulses circulating the microcavity, which corresponds to a coherent optical frequency comb in the spectral domain. Such soliton microcombs present new opportunities of relevance in optics[1,11], such as telecommunications[12], lidar[13,14], optical frequency synthesis[15], microwave photonics[16,17], calibration of astronomical spectrographs[18,19], and quantum optics[20]. The bulk of DKS studies has been conducted using single cavities. In an optical resonator displaying anomalous dispersion, the DKS features a hyperbolic-secant profile[7,21], whose coherence and robustness has been illustrated by multiple demonstrations. In the presence of CW bistability, these waveforms can be formed through modulational instability (MI) by tuning the CW laser from the blue side of resonance towards the red[7,22]. The DKS can only be maintained with the laser effectively red-detuned (see Fig. 1a), where the number of comb lines and generated comb power tends to increase with higher detuning[9,23]. Unfortunately, higher laser detuning reduces the amount of power coupled into the cavity. This imposes a fundamental limitation in the conversion efficiency, i.e. the ratio of power converted from the input CW pump to other frequency components at the output[8].

The low conversion efficiency translates into comb lines with limited power, and places stronger requirements in the performance of on-chip lasers, interposers and frequency doublers for realizing self-referencing[24], a key ingredient in modern frequency synthesis and metrology. Improving the conversion efficiency is instrumental to leverage advances in photonic integration[5,6,25] and to realize fully integrated microcomb-based systems on chip.

The problem of the limited conversion efficiency is exclusive to single DKSs in anomalous dispersion microresonators. Other coherent microcomb states, such as soliton crystals[26,27], Turing rolls[28], or dark-pulse Kerr combs[25,29,30], display much higher conversion efficiency, but they come with other caveats, such as a limitation in the number of lines or narrower bandwidth compared to single DKSs in anomalous-dispersion microresonators. Pulsed pumping improves the conversion efficiency, but this is more cumbersome from an integration perspective, as it requires a comb to generate a DKS microcomb[31]. Another proposal for boosting the conversion efficiency is to use an external cavity with optical gain[32]. However, the high efficiency of this solution has thus far been severely limited with regards to the number of comb lines, and it is unclear if it can operate beyond the laser-gain bandwidth.

Recently, new comb states and soliton dynamics have been discovered in coupled-cavity systems[33–35]. These photonic devices are often termed as "photonic molecules" because their eigenfrequency distribution is akin to the energy levels in molecular and solid-state systems[10,36]. Microcombs generated in photonic diatomic molecules offer a path to high power conversion efficiency[37,38], but the demonstration of DKS microcombs in the anomalous dispersion regime has hitherto remained elusive. Reference [39] demonstrates an arrangement with two linearly coupled cavities, one displaying normal dispersion (where the pump is fed) and another with anomalous dispersion, where the DKS is generated. The intermediate normal-dispersion cavity is described as a storage unit which recycles the pump, enabling high efficiency comb generation. The pump recycling concept was experimentally demonstrated in fiber cavities with pulse initiated DKSs, leading to an increase in soliton energy. However, to the best of our knowledge, high efficiency single DKSs have not been experimentally achieved using this arrangement. Here, we show a different configuration where pump recycling is not strictly required for the enhancement in power conversion efficiency, rather the principal mechanism is the shift induced to the pump resonance of the anomalous soliton cavity (see Fig. 1bi-ii). Our numerical investigations show that by artificially shifting the pump resonance of a *single* anomalous-dispersion cavity, it is possible to operate the soliton with the pump close to center of resonance while other comb lines experience high red-detuning. Physically, this allows the pump to be coupled more efficiently into the cavity while fulfilling the red-detuned criteria of DKSs, resulting in DKS microcombs with a conversion efficiency exceeding 99% (see Fig. 1biii-iv). Because shifts on specific resonances can be attained in other photonic devices, such as photonic crystal resonators[40], we anticipate that high efficiency solitons could be in principle attained in systems different to the photonic molecules discussed in this work. Our results provide a clear pathway towards designing and operating high-efficiency DKSs microcombs, and realizing practical systems that benefit from their exceptional bandwidth and frequency stability.

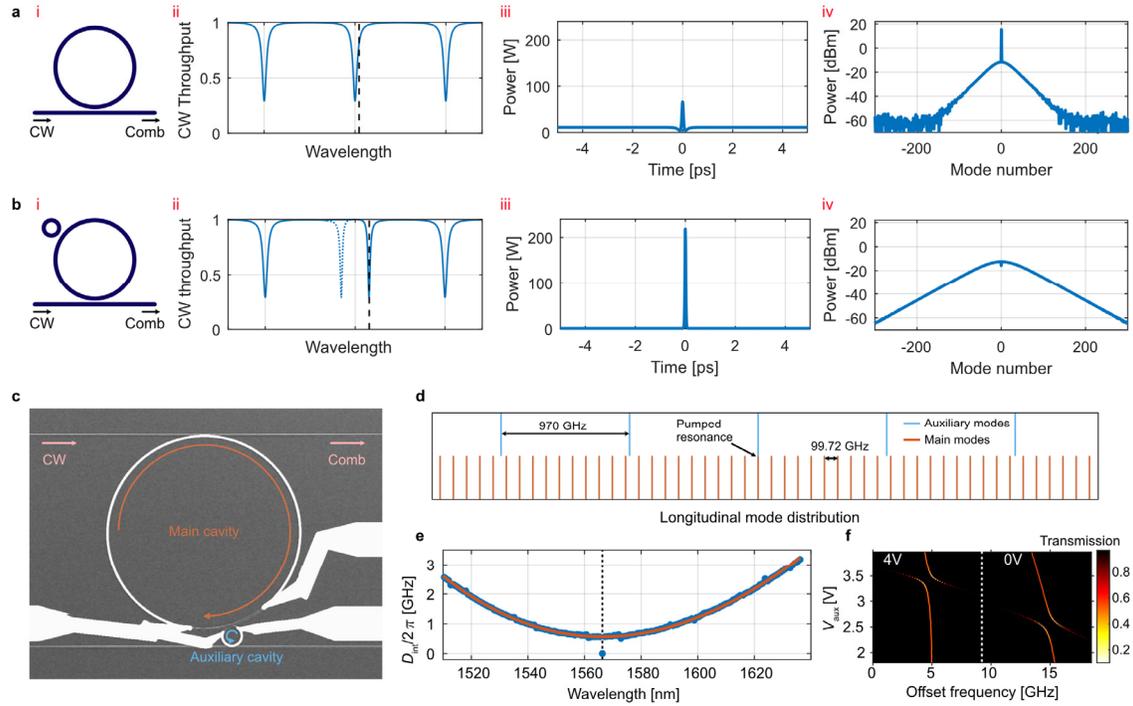

**Fig. 1 Concept of inducing a phase shift to the pump mode for power-efficient comb generation. a,** Comb generation in a single microring resonator (i), whose separation between resonances is defined by the cavity length and the anomalous-dispersion of the waveguides (ii). Using a red-detuned CW laser (dashed line), such configuration can exhibit the generation of a temporal soliton (iii). Much of the CW laser bypasses the cavity, resulting in an output spectrum (iv) with a strong CW component and low comb power[8]. **b,** The same microring now has a mechanism that causes a phase shift *only* to the pump mode, e.g. by inducing a mode-crossing using linear coupling to a small auxiliary cavity (i). This causes the pump resonance to appear shifted while other resonances remain in the same place (ii). The pump can now be shifted further towards the red side, while still coupling ample amount of power into the cavity. This results in higher soliton power (iii) and comb power (iv). **c,** A representative image of a fabricated photonic molecule taken with a scanning electron microscope. The white tracks are metallic heaters. **d** Illustration of the modal distribution of the two cavities, where the mismatch in FSR allows a strong mode-shift to be applied for one longitudinal mode at a time. This is reflected in the dispersion measurement in **e**, showing a relatively smooth anomalous profile of the main cavity with a strong shift at the pump mode (dashed line) induced by coupling to the auxiliary cavity. **f** shows that the mode-crossing induced by the auxiliary ring can be controlled by tuning the auxiliary heater voltage ($V_{aux}$) and through the main heater voltage (set to 0V and 4V).

## Results

### Controllable frequency shift in photonic molecules

We consider a photonic molecule arrangement of two linearly coupled cavities with anomalous dispersion and largely dissimilar volumes (see Fig. 1c-f). The pump is coupled directly into the main cavity while the pump resonance is shifted with the aid of an avoided mode-crossing introduced via coupling to an auxiliary cavity. We fabricated such devices in silicon nitride using a subtractive processing method with heaters on top of the cavities[41]. The heaters allow us to control the location of the avoided mode-crossing and the strength of the coupling[42,43], thus tuning the resonance shift of the pump resonance in a controllable manner. Both cavities have identical cross sections, but the auxiliary cavity is shorter, resulting in a large free spectral range (FSR) which minimizes the interaction between cavities at longitudinal modes other than the pump mode (Fig. 1d). The result is a main cavity with anomalous dispersion, as indicated by the integrated dispersion ($D_{int}$) in Fig. 1e with the auxiliary cavity shifting a single resonance location. The dispersion profile is similar to a previous study of photonic crystal resonator[40], except that our photonic molecule system has the flexibility of introducing a tunable shift (see Fig. 1f). By pumping the shifted resonance with a CW laser, we achieve a power conversion efficiency exceeding 50%, which is an improvement by an order of magnitude compared to previous experimental demonstrations in CW pumped anomalous dispersion microresonators[44]. We further unravel the pathway to the generation of power-efficient microcombs from a CW laser using a bi-dimensional existence map[23,45]. The analysis results in unexpected dynamics, such as the existence of DKSs on both the blue side and the red side of the resonance, the possibility of backwards DKS initiation and DKS operation below MI threshold, properties idiosyncratic to these pump-shifted cavities.

**Mapping the existence of DKSs with a shifted pump resonance**

To investigate the impact of shifting the pump resonance in an anomalous dispersion cavity, we conduct a numerical investigation of the existence map of DKSs in terms of CW pump power and laser detuning. The main objective is to map the conversion efficiency of single-DKSs and the presence modulational instability (MI). Other comb states not corresponding to a single-DKSs (Turing rolls, chaos, multi-DKSs) were also observed, but their characterization is left out for a later study. The dynamics of the system are simulated with an Ikeda map[46], with the pump resonance shift induced by an extra phase shift applied to the pump frequency each roundtrip[29]. Effectively, this leads to two different detuning factors: the detuning between the CW pump frequency and the shifted resonance (defined as *pump detuning*); and the detuning experienced by the other DKS frequencies (defined as *comb detuning*), (see Fig. 2a). Similar to previous investigations of the unaltered anomalous-dispersion cavity[23,45], we use the mean-field approximation to plot the existence map in terms of normalized power ($X$), normalized comb detuning ($\Delta_c$) and normalized pump detuning ($\Delta_p$), (see Methods for definitions).

We begin the analysis by considering a microresonator with zero phase shift applied to the pump (Fig. 2b). Bistability is found in the area confined by the solid line and the dashed line, where the solid line marks the up-switching point[47]. The distribution of comb states across the existence map is in agreement with previous works[45,48]. As expected for standard anomalous-dispersion cavities, the power conversion efficiency is rather low, limited to the range of 4-8%. The absolute maximum conversion efficiency is found to be 8% at a location of roughly 2.5 normalized power with the corresponding DKS waveforms plotted in Fig.1 a. MI is found on the blue side of the CW bistability above a threshold near $X = 1$, as previously predicted[49,50].

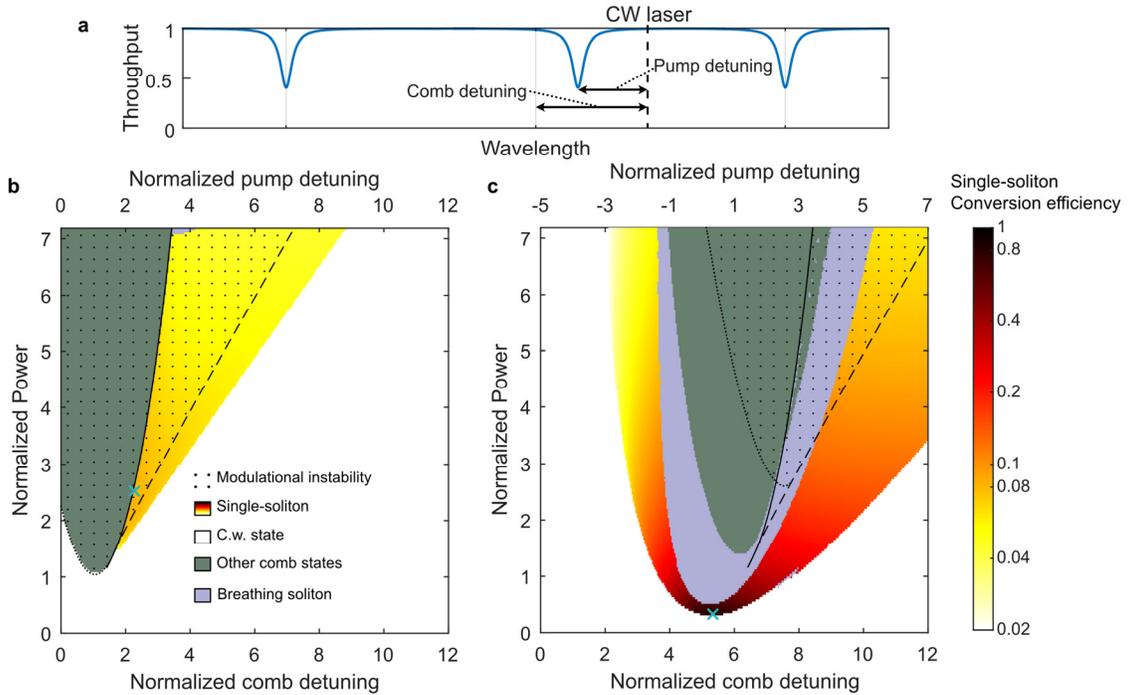

**Fig. 2 Existence maps showing conversion efficiency of single-DKS state. a,** A resonance profile with a shifted center resonance, showing the definition of comb detuning and pump detuning. The two detuning parameters become identical when no shift is applied. **b,** Map for anomalous dispersion cavity without shift applied to the pump resonance. The areas confined by the dashed lines indicate the presence of CW bistability. The dotted areas indicate where combs could be initiated through MI. The peak conversion efficiency is 8% at a point indicated by 'x' in the map. **c,** An existence map where the pump resonance has been shifted by 5 normalized units. The peak conversion efficiency is 80% at the point indicated by 'x' in the map.

In Fig. 2c we use the same cavity, but with the pump resonance shifted by 5 normalized detuning units. Now, single DKSs exist in a continuous area which stretches between the red and blue side of pump resonance, a surprising result considering that DKSs can only exist on the red side of pump resonance in standard anomalous-dispersion cavities. The reason for DKSs existing with a blue detuned pump is the fact that the DKS comb-lines experience red detuning, i.e. the comb detuning always has a positive value. Our study of the closed-form DKS solution shows that the comb detuning determines the shape of the DKS (see Supplementary Information S1). This suggests that the detuning of the DKS is in fact determined by the comb detuning rather than the pump detuning.

The conversion efficiency can now easily reach double digit percentage. Such an enhancement is due to a twofold effect attained by the local frequency shift induced on the pump. On one hand, for a fixed pump power, the frequency shift allows the soliton to exist at much larger comb detuning values compared to the unperturbed cavity. Operating at large comb detuning values enhances the soliton energy and bandwidth[9,23]. The second effect is seen by noting that solitons can exist in a region where the effective pump detuning is zero, so that the pump power is efficiently coupled into resonance and allowing the comb to be operated at a lower input power. The combined effect results in a dramatic improvement in power conversion efficiency (80% for parameters $\Delta_c = 5.32$ and $X = 0.32$), which could in principle result in >99% conversion efficiency if the resonance was shifted further. In the Supplementary Information, we extend the analysis of the high efficiency DKSs using the closed-form solutions of the soliton electric field amplitude.

The single DKS shares a boundary with breather states[51], which were not present at such low power levels in Fig. 2b. Other states appear at the inner boundary of the breather states, which we observe to be mostly chaotic. These can be initiated from MI, near the bistability region of the pump resonance. The MI threshold has been shifted from $X = 1$ of the unperturbed cavity to above $X = 2.5$, which is a consequence of the high comb detuning impeding the phase-matching condition required for MI (see Supplementary Information S2). This absence of MI at lower power levels might be the reason the DKS is allowed to exist at the center of resonance at low power, which is an effect not found in unaltered anomalous-dispersion cavities.

The excitation pathway leading to the formation of DKSs from a CW laser has been radically changed due to the presence of the controllable frequency shift on the pump mode. The fact that the maximum efficiency DKS is located far below the MI threshold of $X = 1$ means that the input power will have to be lowered after initiation. Also, to minimize the initiation power, the initiation should be done with an unshifted pump resonance, with the shift only applied once a comb has been generated. These aspects are utilized as we study the initiation of high efficiency solitons in photonic molecules in the next sections.

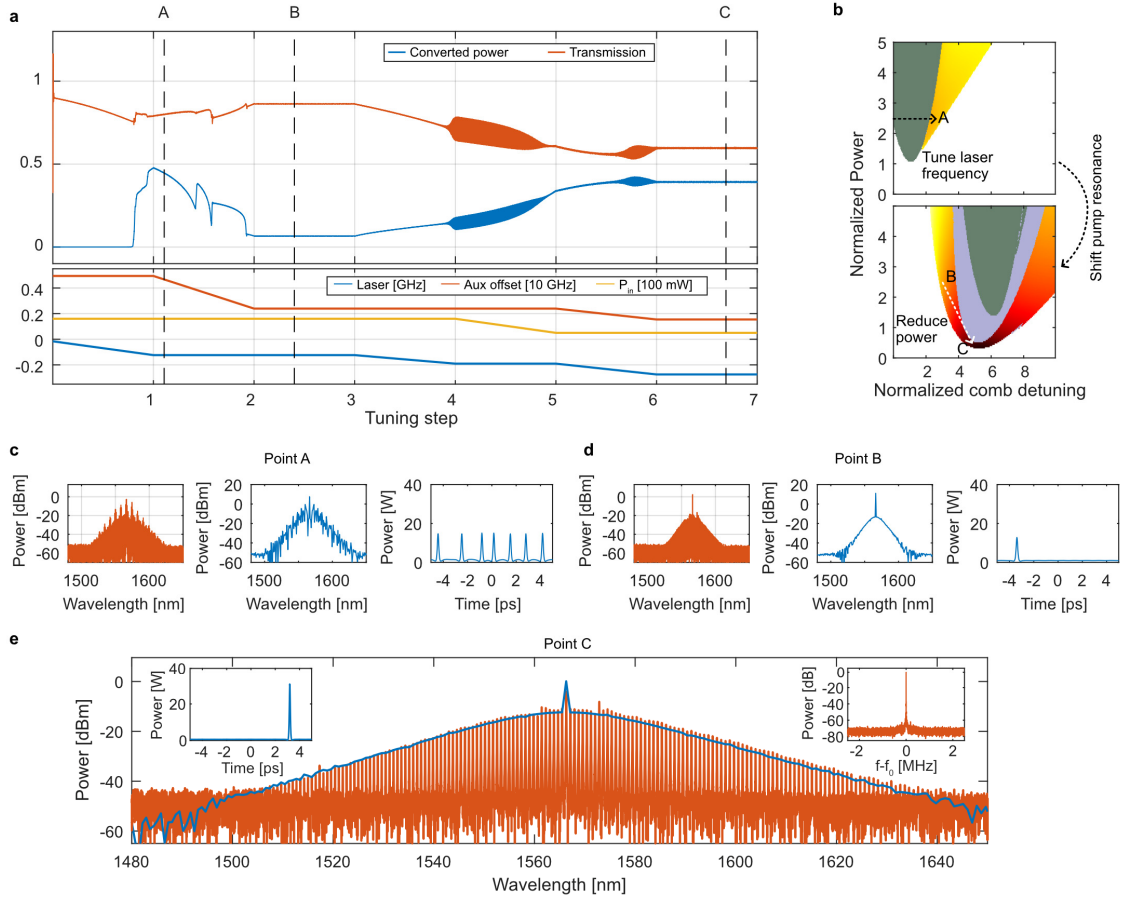

**Fig. 3 Pathway to power-efficient DKS microcombs. a** Simulated initiation process of a single soliton using the photonic molecule configuration. The transmission here is defined as output power divided by input power. The results of the points A-C in the initiation process are indicated by dashed lines, with the corresponding waveforms shown in **c-e**. **b** A qualitative display of the initiation steps required to reach points A-C, using previously presented existence maps. **c** The output spectrum at point A after tuning the laser into resonance from the blue side, both for simulation (blue) and measurement (red). The simulated intracavity temporal field displays multiple solitons in the cavity. **d** The single soliton state acquired after inducing more pump shift by tuning the heater of the auxiliary cavity. This tuning is qualitatively equivalent to moving from point A to point B in the existence map. **e** The soliton state with enhanced conversion efficiency, after reducing the pump power and fine tuning the laser frequency and auxiliary heater. The insets show the intracavity temporal field of the main ring (left) and the measured beat note between comb lines, acquired by electro-optic downconversion and measured with a resolution bandwidth of 100 Hz (right). The final measured spectrum has 51% conversion efficiency.

**From continuous-wave to power-efficient solitons**

To experimentally demonstrate a high efficiency DKS, we use a device with a layout similar to Fig. 1c-d. The integrated dispersion measurement of the main cavity is shown in Fig. 1e. The FSR is 99.72 GHz and group velocity dispersion (GVD) coefficient $\beta_2 = -89\ ps^2/km$. The FSR of the auxiliary cavity was 969.5 GHz. The heaters of the cavities were set such that a resonance split appears near 1566.3nm (see Fig. 1f). The red-shifted hybridized resonance was pumped with a CW laser to generate a comb.

The pathway towards DKS initiation in our experiments involves a series of steps where we slowly tune the laser frequency, auxiliary heater and laser power. To illustrate this pathway, we replicated our DKS initiation in simulation (see Fig. 3a), using a model extended to include coupling to the auxiliary cavity. The settings of the simulation are based on measured characteristics of the device. The change in comb states is seen in the variation of converted and transmitted power. A qualitative depiction of the initiation process is shown in Fig. 3b. Fig. 3c-e shows the microcomb states at the indicated points in the initiation pathway, both from experiments and simulation. We start with the pump resonance almost decoupled from the auxiliary resonance, exploiting the fact that MI can be attained at lower pump powers in single cavity systems. In practice, the laser starts blue detuned from resonance with about 16mW of on-chip CW input power. The auxiliary resonance was placed roughly 5 GHz away towards the blue side of the main resonance, corresponding to an auxiliary voltage of approximately 2.05 V and a main cavity voltage of 0 V. We first move towards point A by changing the voltage applied to the piezo element in the laser, slowly tuning the laser frequency into resonance until a multi-DKS state was generated (see Fig. 3c). We then tune to point B by increasing the voltage of the auxiliary heater, thereby increasing the shift induced to the pump resonance. The number of solitons is decreased step by step, until a single soliton state is attained. The dynamics are captured by our simulations, resulting in a low power DKS (see Fig. 3d), in agreement with the measured soliton spectrum. Once a single DKS is obtained in the photonic molecule, the strategy to reach high conversion efficiency is to adiabatically modify the detuning and pump power in order to reach the area at low pump powers where the normalized pump detuning is close to zero in Fig. 2c. We tuned the power down to 5mW while fine tuning the auxiliary heater voltage and the laser piezo. This is done stepwise in our simulations, revealing a transition through the breather region before reaching the point C. The final DKS comb state is shown in Fig. 3e and it features a smooth sech spectrum, exhibiting 51% conversion efficiency in experimental measurement. The inset shows a measurement of the comb's repetition rate measured through electro-optic down-conversion[52], indicating a coherent comb state. The simulated temporal field indicates that the spectrum corresponds to a single DKS circulating the cavity. It is worth emphasizing that the 16mW initiation power used here was selected for ease of operation, as it allowed the comb to be initiated by simply tuning one instrument at a time by hand. Initiation at a lower power level is more sensitive and requires more elaborate schemes such as feedback locking between comb power and laser frequency[53]. We found that such feedback locking could be employed to initiate the comb at 8-9mW of input power.

In the supplementary information, we numerically show how the DKS can be scaled both in power and in number of lines generated. Furthermore, using another device, we experimentally demonstrate that by increasing the coupling rate between bus and main cavity we can increase the comb power while reaching a conversion efficiency of 55%.

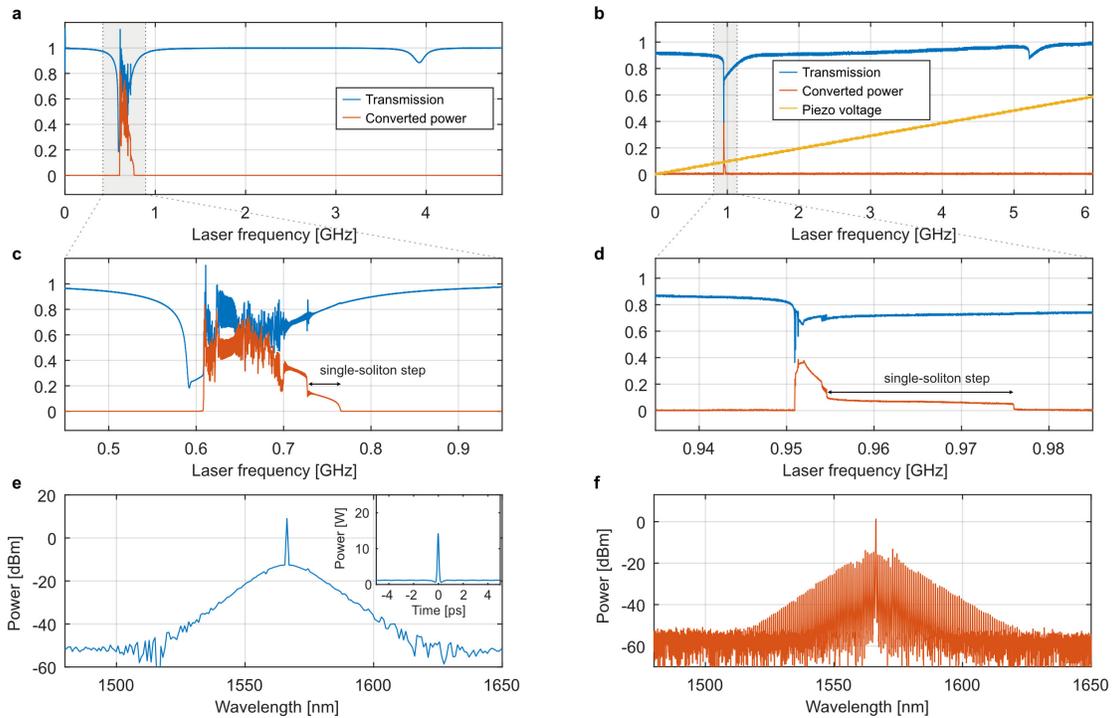

**Fig. 4 Backwards initiation of solitons. a** A simulated scan of laser frequency, tuning across the coupled resonances from the red side, i.e. starting from the left and moving towards the right. The scan reveals the main resonance (left side) which is slightly shifted due to the proximity of an auxiliary resonance (right side). The converted power rises in the main resonance where a comb is generated. **b** shows a measurement using the same configuration. The laser frequency is swept up by increasing the piezo voltage of the laser. **c-d** A close-up of the simulated and measured laser scan, revealing a soliton step. **e-f** The simulated and measured soliton acquired when stopping the laser on the soliton step. The inset in **e** shows the main-cavity temporal field.

**Backwards initiation and blue side operation of DKS states**

The existence map in Fig. 2c shows that the single-DKS states exist not only on the red side, but also on the blue side of the shifted pump resonance. Such blue-detuned operation is highly interesting because in standard anomalous dispersion cavities the DKS can only operate with the CW laser effectively on the red side of resonance[22]. Furthermore, with high enough input power, Fig. 2c anticipates that these blue-detuned DKSs exist on the blue side of MI, indicating that the DKSs can be initiated by tuning the CW laser into resonance from the red side towards the blue, another phenomenon not found in unperturbed anomalous-dispersion cavities. We perform two sets of experiments to verify these dynamics using the same device and simulation model as in Fig.3. The first measurement investigates the red-side initiation of DKSs with the results displayed in Fig. 4 along with simulations. Figure 4a-b show both simulated and measured oscilloscope scans of the transmission and converted comb power as the laser is swept up in frequency, starting on the red side of resonance. The auxiliary heater is kept at a fixed level and the on-chip input power is estimated to be 11.5 mW. The transmission trace displays the main resonance on the red side which is influenced by the proximity of the auxiliary resonance. The trace of converted comb power indicates a comb is generated when pumping the shifted main resonance. A close inspection of Fig. 4c-d shows that the converted trace exhibits soliton steps, with good agreement between simulation and measurement. Stopping the laser on the soliton step, we encounter a single-DKS state (see Fig.4e-f). This state is similar to point B in Fig. 3d, where the same type of tuning as displayed in Fig.3a can be used to reach a high efficiency DKS.

Operating the DKS at 5 mW (similar to Fig. 3e), we conduct our second experiment to find if the pump is blue-detuned or red-detuned when generating a DKS. For this we probe the cavity with a counter-propagating probe-laser[54]. Meanwhile, the pump is operated using a feedback lock between laser frequency and generated comb power to ensure long-term stability, which was essential to maintain the comb with the laser red detuned. The transmission of the probe as a function of frequency is displayed in Fig. 5a and the corresponding comb state shown in Fig. 5b. The probe trace shows the two hybridized resonances with the pumped resonance located on the red side. The pumped resonance exhibits a clear beat note between the probe laser and pump laser appearing on the blue side of resonance, indicating that the comb is indeed operated with a blue-detuned laser.

When the pump frequency is tuned towards the red side of the resonance, it eventually results in the beat note flipping to the red side of resonance with a slight change in the spectral distribution of the comb (see Fig.5c-d). This suggests that we are operating at a point similar to the bottom of the DKS existence in Fig. 2c, where blue-detuned and red-detuned DKSs exist in a continuous stretch. This is to our knowledge the first time a soliton in the anomalous dispersion regime is operated effectively blue-detuned to pump resonance.

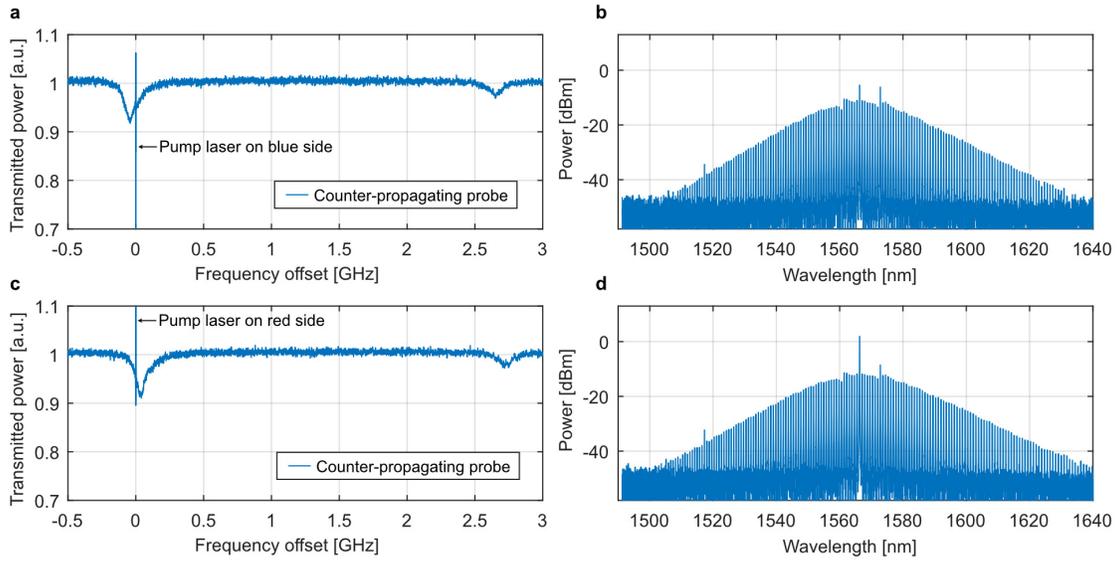

**Fig. 5 Microcombs operated on blue or red side of resonance. a** The response of the counter-propagating probe-laser, showing that the beating with the pump laser occurs on the blue side of cavity resonance. **b** The power spectral distribution of the corresponding comb measured in an optical spectrum analyzer (OSA). **c** The response of the counter-propagating probe-laser after shifting the pump frequency to the red side of the resonance. **d** shows the corresponding comb state as measured in the OSA.

## Highly coherent 50 GHz DKS

In this section, we demonstrate that highly efficient DKS microcombs can be operated with different cavity volumes. For this, we double the length of both the main- and auxiliary cavity resulting in an FSR of 49.85 GHz and 486.5 GHz, respectively. The DKS microcomb operates at a repetition rate commensurate with electronics bandwidth, which is relevant for applications in telecommunications[30,55,56], radio-frequency photonics[17] and the calibration of astronomical spectrographs[18]. Enhancing the cavity volume is challenging with planar integrated technologies, and the decrease associated with intracavity intensity must encompassed with an increase in quality factor and pump power[57]. The large conversion efficiency allows us to relax these requirements while maintaining outstanding purity in repetition rate frequency. A simplified schematic of the setup is displayed in Fig. 6a, which also shows how the device was optically coupled using lensed fibers. The heaters were tuned such that the auxiliary cavity induced a resonance split near 1550.1 nm. This split resonance was pumped with a low phase-noise fiber-laser at a fixed wavelength of 1550.1 nm. The output spectrum of the comb was characterized in an OSA (see Fig. 6b), resulting in a DKS with 35% conversion efficiency. The repetition rate of the comb was measured directly as a beatnote using a high bandwidth photodiode connected to an electrical spectrum analyzer (ESA). The result was a narrow tone located near 49.85 GHz with low power spectral density (see Fig. 6c), indicating a coherent comb state with a phase noise comparable to state-of-the-art DKS microcombs[16].

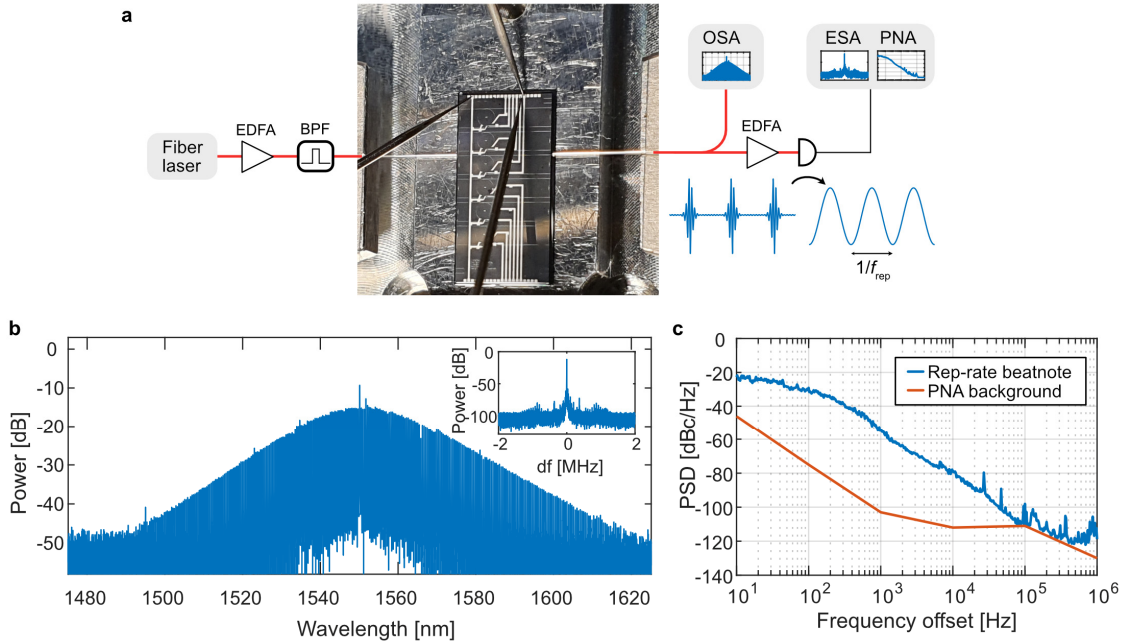

**Fig. 6 Highly efficient soliton microcombs and microwave frequency synthesis. a** A simplified diagram of the measurement setup, including a picture of the optically coupled chip with power supplied to heater pads through positional probes. A high-speed photodiode transforms the output pulses into a microwave tone at a frequency equal to the repetition rate. **b** The measured spectrum of a microcomb operated with 35% conversion efficiency. The inset shows the repetition-rate beatnote at 49.85 GHz measured in an electrical spectrum analyzer (ESA). **c** The power spectral density (PSD) of the repetition-rate beatnote measured using a phase noise analyzer (PNA).

**Discussion**

We have demonstrated high efficiency microcombs generated in anomalous dispersion microresonators, enabled by shifting the CW pumped resonance of the microcavity. This shift was realized in practice using an avoided mode-crossing, induced by an auxiliary resonator coupled to the main cavity. We experimentally demonstrated three devices operating with different input powers, each exhibiting a smooth DKS spectrum with up to 55% conversion efficiency. To our knowledge, this is the highest conversion efficiency achieved for a single soliton state generate in a microresonator from a CW pump. This high conversion efficiency and uniform spectrum make these devices highly relevant for applications in optical communications[12] and dual-comb spectroscopy[58]. With further engineering of the coupling region and dispersion, we believe these results pave the way for the realization of octave-spanning microcombs and self-referencing using only integrated components.

**Data availability**

The raw data will be published on zenodo.org upon publication of this work.

**References**

1. Diddams, S. A., Vahala, K. & Udem, T. Optical frequency combs: Coherently uniting the electromagnetic spectrum. *Science (80-. ).* **369**, eaay3676 (2020).

2. Fortier, T. & Baumann, E. 20 years of developments in optical frequency comb technology and applications. *Commun. Phys.* **2**, 153 (2019).

3. Kippenberg, T. J., Gaeta, A. L., Lipson, M. & Gorodetsky, M. L. Dissipative Kerr solitons in optical microresonators. *Science (80-. ).* **361**, eaan8083 (2018).

4. Stern, B., Ji, X., Okawachi, Y., Gaeta, A. L. & Lipson, M. Battery-operated integrated frequency comb generator. *Nature* **562**, 401–405 (2018).

5. Xiang, C. *et al.* Laser soliton microcombs heterogeneously integrated on silicon. *Science (80-. ).* **373**, 99–103 (2021).

6. Shen, B. *et al.* Integrated turnkey soliton microcombs. *Nature* **582**, 365–369 (2020).

7. Herr, T. *et al.* Temporal solitons in optical microresonators. *Nat. Photonics* **8**, 145–152 (2014).

8. Xue, X., Wang, P.-H., Xuan, Y., Qi, M. & Weiner, A. M. Microresonator Kerr frequency combs with high conversion efficiency. *Laser Photon. Rev.* **11**, 1600276 (2017).

9. Bao, C. *et al.* Nonlinear conversion efficiency in Kerr frequency comb generation. *Opt. Lett.* **39**,

6126–6129 (2014).

10. Zhang, M. *et al.* Electronically programmable photonic molecule. *Nat. Photonics* **13**, 36–40 (2019).

11. Pasquazi, A. *et al.* Micro-combs: A novel generation of optical sources. *Phys. Rep.* **729**, 1–81 (2018).

12. Marin-Palomo, P. *et al.* Microresonator-based solitons for massively parallel coherent optical communications. *Nature* **546**, 274–279 (2017).

13. Suh, M.-G. & Vahala, K. J. Soliton microcomb range measurement. *Science (80-. ).* **359**, 884–887 (2018).

14. Trocha, P. *et al.* Ultrafast optical ranging using microresonator soliton frequency combs. *Science (80-. ).* **359**, 887–891 (2018).

15. Spencer, D. T. *et al.* An optical-frequency synthesizer using integrated photonics. *Nature* **557**, 81–85 (2018).

16. Liu, J. *et al.* Photonic microwave generation in the X- and K-band using integrated soliton microcombs. *Nat. Photonics* **14**, 486–491 (2020).

17. Wu, J. *et al.* RF Photonics: An Optical Microcombs' Perspective. *IEEE J. Sel. Top. Quantum Electron.* **24**, 1–20 (2018).

18. Suh, M.-G. *et al.* Searching for exoplanets using a microresonator astrocomb. *Nat. Photonics* **13**, 25–30 (2019).

19. Obrzud, E. *et al.* A microphotonic astrocomb. *Nat. Photonics* **13**, 31–35 (2019).

20. Guidry, M. A., Lukin, D. M., Yang, K. Y., Trivedi, R. & Vučković, J. Quantum optics of soliton microcombs. *Nat. Photonics* **16**, 52–58 (2022).

21. Leo, F. *et al.* Temporal cavity solitons in one-dimensional Kerr media as bits in an all-optical buffer. *Nat. Photonics* **4**, 471–476 (2010).

22. Guo, H. *et al.* Universal dynamics and deterministic switching of dissipative Kerr solitons in optical microresonators. *Nat. Phys.* **13**, 94–102 (2017).

23. Jaramillo-Villegas, J. A., Xue, X., Wang, P.-H., Leaird, D. E. & Weiner, A. M. Deterministic single soliton generation and compression in microring resonators avoiding the chaotic region. *Opt. Express* **23**, 9618–9626 (2015).

24. Rao, A. *et al.* Towards integrated photonic interposers for processing octave-spanning

microresonator frequency combs. *Light Sci. Appl.* **10**, 109 (2021).

25.	Jin, W. *et al.* Hertz-linewidth semiconductor lasers using CMOS-ready ultra-high-Q microresonators. *Nat. Photonics* **15**, 346–353 (2021).

26.	Cole, D. C., Lamb, E. S., Del'Haye, P., Diddams, S. A. & Papp, S. B. Soliton crystals in Kerr resonators. *Nat. Photonics* **11**, 671–676 (2017).

27.	Corcoran, B. *et al.* Ultra-dense optical data transmission over standard fibre with a single chip source. *Nat. Commun.* **11**, 2568 (2020).

28.	Coillet, A. *et al.* Azimuthal Turing Patterns, Bright and Dark Cavity Solitons in Kerr Combs Generated With Whispering-Gallery-Mode Resonators. *IEEE Photonics J.* **5**, 6100409 (2013).

29.	Xue, X. *et al.* Mode-locked dark pulse Kerr combs in normal-dispersion microresonators. *Nat. Photonics* **9**, 594–600 (2015).

30.	Fülöp, A. *et al.* High-order coherent communications using mode-locked dark-pulse Kerr combs from microresonators. *Nat. Commun.* **9**, 1598 (2018).

31.	Anderson, M. H. *et al.* Photonic chip-based resonant supercontinuum via pulse-driven Kerr microresonator solitons. *Optica* **8**, 771 (2021).

32.	Bao, H. *et al.* Laser cavity-soliton microcombs. *Nat. Photonics* **13**, 384–389 (2019).

33.	Mittal, S., Moille, G., Srinivasan, K., Chembo, Y. K. & Hafezi, M. Topological frequency combs and nested temporal solitons. *Nat. Phys.* **17**, 1169–1176 (2021).

34.	Marti, L., Vasco, J. P. & Savona, V. Slow-light enhanced frequency combs and dissipative Kerr solitons in silicon coupled-ring microresonators in the telecom band. *OSA Contin.* **4**, 1247 (2021).

35.	Tikan, A. *et al.* Emergent Nonlinear Phenomena in a Driven Dissipative Photonic Dimer. *Nat. Phys.* **17**, 604–610 (2020).

36.	Bayer, M. *et al.* Optical Modes in Photonic Molecules. *Phys. Rev. Lett.* **81**, 2582–2585 (1998).

37.	Helgason, Ó. B. *et al.* Dissipative solitons in photonic molecules. *Nat. Photonics* **15**, 305–310 (2021).

38.	Kim, B. Y. *et al.* Turn-key, high-efficiency Kerr comb source. *Opt. Lett.* **44**, 4475–4478 (2019).

39.	Xue, X., Zheng, X. & Zhou, B. Super-efficient temporal solitons in mutually coupled optical cavities. *Nat. Photonics* **13**, 616–622 (2019).

40.	Yu, S.-P. *et al.* Spontaneous pulse formation in edgeless photonic crystal resonators. *Nat.*

*Photonics* **15**, 461–467 (2021).

41.  Ye, Z., Twayana, K., Andrekson, P. A. & Torres-Company, V. High-Q Si 3 N 4 microresonators based on a subtractive processing for Kerr nonlinear optics. *Opt. Express* **27**, 35719 (2019).

42.  Xue, X. *et al.* Normal-dispersion microcombs enabled by controllable mode interactions. *Laser Photon. Rev.* **9**, L23–L28 (2015).

43.  Gentry, C. M., Zeng, X. & Popović, M. A. Tunable coupled-mode dispersion compensation and its application to on-chip resonant four-wave mixing. *Opt. Lett.* **39**, 5689 (2014).

44.  Liu, J. *et al.* Ultralow-power chip-based soliton microcombs for photonic integration. *Optica* **5**, 1347–1353 (2018).

45.  Godey, C., Balakireva, I. V, Coillet, A. & Chembo, Y. K. Stability analysis of the spatiotemporal Lugiato-Lefever model for Kerr optical frequency combs in the anomalous and normal dispersion regimes. *Phys. Rev. A* **89**, 063814 (2014).

46.  Hansson, T. & Wabnitz, S. Dynamics of microresonator frequency comb generation: models and stability. *Nanophotonics* **5**, 231–243 (2016).

47.  Coen, S. & Erkintalo, M. Universal scaling laws of Kerr frequency combs. *Opt. Lett.* **38**, 1790–1792 (2013).

48.  Barashenkov, I. V & Smirnov, Y. S. Existence and stability chart for the ac-driven, damped nonlinear Schrödinger solitons. *Phys. Rev. E* **54**, 5707–5725 (1996).

49.  Haelterman, M., Trillo, S. & Wabnitz, S. Additive-modulation-instability ring laser in the normal dispersion regime of a fiber. *Opt. Lett.* **17**, 745 (1992).

50.  Haelterman, M., Trillo, S. & Wabnitz, S. Dissipative modulation instability in a nonlinear dispersive ring cavity. *Opt. Commun.* **91**, 401–407 (1992).

51.  Bao, C. *et al.* Observation of Fermi-Pasta-Ulam Recurrence Induced by Breather Solitons in an Optical Microresonator. *Phys. Rev. Lett.* **117**, 163901 (2016).

52.  Del'Haye, P., Papp, S. B. & Diddams, S. A. Hybrid Electro-Optically Modulated Microcombs. *Phys. Rev. Lett.* **109**, 263901 (2012).

53.  Yi, X., Yang, Q.-F., Youl Yang, K. & Vahala, K. Active capture and stabilization of temporal solitons in microresonators. *Opt. Lett.* **41**, 2037 (2016).

54.  Del'Haye, P. *et al.* Phase steps and resonator detuning measurements in microresonator frequency combs. *Nat. Commun.* **6**, 5668 (2015).

55.    Liao, P. *et al.* Dependence of a microresonator Kerr frequency comb on the pump linewidth. *Opt. Lett.* **42**, 779 (2017).

56.    Mazur, M. *et al.* High Spectral Efficiency Coherent Superchannel Transmission With Soliton Microcombs. *J. Light. Technol.* **39**, 4367–4373 (2021).

57.    Ye, Z. *et al.* Integrated, Ultra-Compact High-Q Silicon Nitride Microresonators for Low-Repetition-Rate Soliton Microcombs. *Laser Photon. Rev.* 2100147 (2021) doi:10.1002/lpor.202100147.

58.    Suh, M.-G., Yang, Q.-F., Yang, K. Y., Yi, X. & Vahala, K. J. Microresonator soliton dual-comb spectroscopy. *Science (80-. ).* **354**, 600–603 (2016).

59.    Del'Haye, P., Arcizet, O., Gorodetsky, M. L., Holzwarth, R. & Kippenberg, T. J. Frequency comb assisted diode laser spectroscopy for measurement of microcavity dispersion. *Nat. Photonics* **3**, 529–533 (2009).

**Acknowledgements**

The simulations for Fig. 2 were performed on resources at Rackham provided by the Swedish National Infrastructure for Computing (SNIC). The devices demonstrated in this work were fabricated in part at Myfab Chalmers. We acknowledge funding support from: European Research Council (ERC, CoG GA 771410); Swedish Research Council (2016-06077, 2020-00453, 2017-05157).

## Materials and methods

### Device characterization

The devices were fabricated on a chip using a silicon nitride waveguide core, with a silica cladding using a subtractive processing method[41]. Heaters were placed on both auxiliary and main cavity, using positional probes to connect voltage to the heaters. The devices were accessed by coupling light into the bus waveguide at the chip facets using a lensed fiber. Most of the devices were spectrally characterized using a self-referenced mode-locked laser as a reference[59], in a manner similar to our previous work[37]. Other measurements featured spectral characterization using a calibrated fiber interferometer. Device 1, used in the experimental demonstration of Fig. 3, Fig. 4 and Fig.5 had a facet coupling loss of ~3dB per facet. It had a main cavity of radius 227.82 µm and the radius of the auxiliary cavity was 23.32 µm. All waveguides had the same dimensions of 1800 nm width and 740 nm height. The gap between rings was 500nm, the gap between main cavity and bus was 400nm, and the gap between auxiliary cavity and bus was 700 nm. The main cavity was measured (referenced to 1566 nm) with GVD of $\beta_2 = -89\ ps^2/km$, FSR of 99.72 GHz, intrinsic Q factor with an average near 8 million, and a coupling rate to bus waveguide corresponding to extrinsic Q of 2-3 million near 1566nm. The maximal resonance shift induced by the auxiliary cavity was measured as 770 MHz. The auxiliary cavity resonance at 1566.3 nm which interacted with the main cavity had 1.2 million intrinsic Q and coupling rate to a bus waveguide corresponding to 50 million of extrinsic Q. The auxiliary FSR was 969.5 GHz.

Device 2, used in the experimental demonstration of Fig. 6 had a facet coupling loss of ~2dB per facet. It had a main cavity of radius 455.02 μm and the radius of the auxiliary cavity was 46.62 μm. All waveguides had the same dimensions of 1800 nm width and 740 nm height. The gap between rings was 450nm, the gap between main cavity and bus was 350nm, and the gap between auxiliary cavity and bus was 700 nm. The main cavity was measured (referenced to 1550 nm) with GVD of $\beta_2 = -82\ ps^2/km$, FSR of 49.86 GHz, intrinsic Q factor with an average near 10 million, and a coupling rate to bus waveguide corresponding to extrinsic Q of 2-3 million near 1550nm. The maximal resonance shift induced by the auxiliary cavity was measured as 440 MHz. The auxiliary cavity resonance at 1550 which interacted with the main cavity had near 4 million intrinsic Q and coupling rate to a bus waveguide corresponding to 60 million of extrinsic Q. The auxiliary FSR was 486.5 GHz. Using a mode-solver for the cross-section of the microresonator waveguides, the nonlinear parameter was estimated to be $\gamma = 0.9\ (W\ m)^{-1}$ for both devices.

The microcomb characterization of spectral distribution, conversion efficiency and repetition rate was conducted using the methods described in our previous work[37], unless otherwise specified.

The combs in Fig.3-5 were pumped with an external cavity diode laser (ECDL) which was amplified with an EDFA. The counter-propagating laser used in Fig. 5 was another ECDL operated at -10 dBm. The trace in Fig. 5 measured in the oscilloscope was low-pass filtered with a 0.5 MHz bandwidth such that a beatnote would only appear near the pump laser frequency. The trace was scaled to remove a constant background coming from the reflected pump.

The measurement in Fig. 6 involved a low phase-noise fiber-laser (NKT Koheras Basik), which operates at a wavelength of 1550.1 nm. The laser was amplified with an EDFA, with the amplified spontaneous emission noise removed in a notch filter just before being coupled into the chip. A comb was initiated using a method similar to that of Fig.3, only the detuning between laser and resonances was managed by changing the cavity heaters with the help of feedback locking using the comb power.

An EDFA was used as a pre-booster for the beatnote detection. The beatnote was amplified in a highspeed electrical amplifier prior to detection in the ESA. The ESA was a PXA signal analyzer N9030A, which has phase noise analyzing functionality to measure the single-sideband PSD.

**Numerical models**

The numerical simulation used in Fig. 1-2 is based on an Ikeda map modified to include a phase shift to the pump frequency. Each roundtrip simulated for the optical field of the microring includes a coupling step, nonlinear propagation around the full length of the microring (L) and a phase shift applied only to the pump frequency. The coupling regime describes the coupling between a straight bus waveguide and a ring waveguide. It is approximated as point coupling given by the following equation:

$$\begin{bmatrix} A_{\text{out}} \\ A \end{bmatrix} = \begin{bmatrix} \sqrt{1-\theta} & \text{i}\sqrt{\theta} \\ \text{i}\sqrt{\theta} & \sqrt{1-\theta} \end{bmatrix} \begin{bmatrix} A_{\text{in}} \\ A' \end{bmatrix},$$

Where A is the temporal field of the ring cavity, $A_{\text{in}}$ is the input pump field in the bus waveguide, $A_{\text{out}}$ is the output field in the bus waveguide and $\theta$ is the portion of power coupled between the bus and ring. The nonlinear propagation is described by the nonlinear Schrödinger equation

$$\left( \frac{\partial}{\partial z} + \frac{\alpha_i}{2} + \text{i}\varphi + \text{i}\frac{\beta_2}{2}\frac{\partial^2}{\partial t^2} - \text{i}\gamma \left| A \right|^2 \right) A = 0,$$

Where z is location in the waveguide, $\alpha_i$ is the propagation loss, $\varphi L$ is the detuning per roundtrip, $\beta_2$ is the group velocity dispersion, t is the reference time of the cavity and $\gamma$ is the nonlinear coefficient. To capture the shift applied to the pump resonance, we modify the detuning parameter according to $\varphi = \varphi_0 + \sigma\delta(\mu)$, where $\varphi_0$ is the detuning of the laser from the unshifted pump resonance (comb detuning), $\sigma$ is the shift factor applied to the pump, $\delta$ is the Dirac delta function and $\mu$ is the frequency mode with $\mu = 0$ corresponding to the pump mode.

The normalization for Fig. 2 is done in the same way as in previous works[37,47]. Concretely, the normalized comb detuning is $\Delta_c = \varphi_0 L/\alpha$, the normalized pump detuning is $\Delta_p = (\varphi_0 + \sigma)L/\alpha$ and the normalized power is $X = P_{in}L\gamma\theta/\alpha^3$, where $\alpha = (\alpha_i L + \theta)/2$. The intracavity field and its temporal distribution are also normalized according to $F = A\sqrt{\frac{\gamma L}{\alpha}}$ and $t' = t\sqrt{\frac{2\alpha}{|\beta_2|L}}$, respectively. Note that by changing $P_{in}$ one can scale the comb to different power levels without change in shape as long as

other parameters are adjusted such that $\Delta_0$, $\Delta_p$, $X$ and $t'$ remain constant. The converted power in Fig. 2 will remain unchanged with such scaling, unless intrinsic losses are introduced. More details on the existence map simulation are provided in the next method section.

The numerical model used for Figs. 3 and 4 was based on the coupled cavity model[37], which captures the nonlinear propagation in both cavities. To mitigate the mismatch in size between cavities, the auxiliary cavity was scaled to have the same length as the main cavity ($2\pi \times 227.82$ μm), resulting in a 10 times smaller FSR (97 GHz). By only allowing every 10th comb line to acquire power, we effectively have a spectral distribution of 970 GHz. The coupling between main cavity and auxiliary cavity was only allowed for every 10th mode, including the pump mode. The main cavity of the simulation had an intrinsic quality factor of 10 million, extrinsic quality factor of 2.5 million, GVD of $\beta_2 = -89\ ps^2/km$ and FSR of 99.72 GHz. The second cavity had quality factor of 1.2 million and dispersion of $\beta_2 = -89\ ps^2/km$. We observed that the dispersion of the second cavity had virtually no impact on our simulations. The nonlinear parameter of both cavities was set to $\gamma = 0.9\ (W\ m)^{-1}$. The ratio of power coupled between the two cavities each roundtrip was 0.24%. The cold-cavity detuning of the laser, and the relative cold-cavity shift between auxiliary resonance and main resonance are provided in Fig. 3a. Note that these detuning values do not include the effects of nonlinearity and resonance split. In Fig. 4b, the laser frequency describes detuning from main resonance with 873 MHz added. The auxiliary resonance was detuned 2.856 GHz towards the blue side of the main resonance.

**Generating existence maps**

The existence maps were generated on a grid of 0.04 resolution of normalized detuning and normalized power. At each grid-point point, the intracavity waveform was propagated up to 80 000 roundtrips so that it would reach a steady-state. After that, 10 000 roundtrips were recorded to characterize the state. Stability and oscillations were tested by measuring of intracavity mean power across these roundtrips. The comb states were acquired in 3 different stages. First the DKS existence was mapped. A single DKS was initiated at some point in the map. The detuning and power were gradually changed to move between grid-points, where the comb state was characterized. The conversion efficiency was defined as the power of output comb lines excluding the pump frequency divided by the input pump power.

A second stage was to observe the presence of modulational instability. For this, the intracavity field was set to a CW steady state of the cavity (selecting the upper branch in presence of bistability) adding a noise seed of low amplitude. These CW steady state solutions were found by adapting previously known equations for single cavities[50], written as $X = Y^3 - 2\Delta_p Y^2 + (\Delta_p^2 + 1)Y$, where $X$ is the normalized input power and $Y$ is the normalized intracavity power and $\Delta_p$ is the normalized pump detuning. The areas which displayed converted power were considered to show MI. In the final stage, starting with intracavity waveforms found at stage 2 as a seed, we tuned to different points of the grid to find potential comb states that might have been missed in the first two stages.

# Power-efficient soliton microcombs – Supplementary Information

Óskar B. Helgason, Marcello Girardi, Zhichao Ye, Fuchuan Lei, Jochen Schröder and

Victor Torres-Company*

Department of Microtechnology and Nanoscience, Chalmers University of Technology, SE-41296

Gothenburg, Sweden

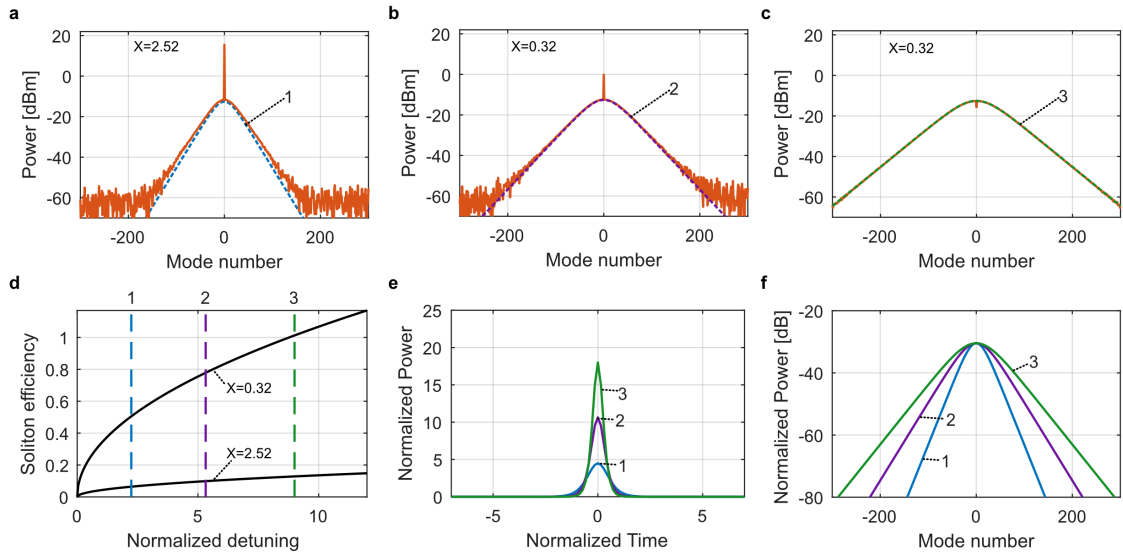

**Fig. S1 Analyzing DKS efficiency using closed form solutions. a-c** The outcoupled spectral distribution of the closed-form solution for detuning 1-3 after transferring to real parameters (dashed), each figure indicating the set power. The red trace in **a** shows the simulation from the standard anomalous cavity in Fig.1a-iv which used the same parameters as the closed form solution. The red trace in **b** shows the simulation from the highest efficiency DKS of the anomalous cavity in Fig.2 b, where the pump resonance was shifted by 5 normalized detuning units. The comb detuning in the simulation matches the detuning of the closed form solution. The red trace in **c** shows a simulation from the shifted anomalous cavity in Fig.1b-iv . The comb detuning in the simulation matches the detuning of the closed form solution. **d**, The change in soliton efficiency according to the closed form solution as a function of detuning for two different levels of normalized Power (X). The efficiency scales with the square root of normalized detuning. The dashed lines mark the detuning of the DKSs in **a-c**. **e** The intracavity field of the soliton for the three detuning levels marked by the dashed lines in **d**. **f** The spectral distribution of the solitons in **e**, showing that the increase in detuning causes the spectrum to get broader.

## S1. Analyzing soliton envelope

The high conversion efficiency of DKSs is enabled by high comb detuning and the ability to operate the pump at the center of resonance with lower power. Here, we show these dynamics are predicted by the closed-analytical solution for the electric field of the cavity soliton[1,2]. This analytical approximation is written as

$$\mathrm{F} = \sqrt{2\Delta}\, \mathrm{sech}\left(\sqrt{\Delta}\, t'\right), \qquad (1)$$

where $\Delta$ is the normalized detuning, $t'$ is the normalized temporal distribution and $\mathrm{F}$ is the normalized intracavity field. These normalized parameters are described further in the method section of the main manuscript. This solution assumes a conservative system, thus it does not depend on power coupled into the cavity or damping factor. It does nonetheless give a good approximation for soliton waveforms in driven-damped microcavities[3]. As an example, in Fig. S1a, we show a good agreement between the Eq. 1 and the simulation of the outcoupled spectrum from Fig. 1a-iv when using the same parameters.

We find that Eq. 1 can be applied to a cavity with a shifted pump resonance by applying the comb detuning parameter (i.e. by setting $\Delta = \Delta_c$). This is demonstrated in Fig. S1b using the settings of the highest efficiency DKS in Fig. 2b, where the main resonance had been shifted by 5 normalized units. The result shows a remarkable agreement between Eq. 1 and simulation, showing that the DKS shape is determined by the comb detuning. We evaluate the results again in Fig. S1c using the same setting as in Fig. 1b-iv, operating the comb with comb detuning $\Delta_c = 9$, pump detuning $\Delta_p = 0.25$ and normalized power $X = 0.32$, resulting in >99% conversion efficiency.

Having established the closed-form expression in Eq. 1 as a useful qualitative approximation for DKSs in anomalous cavities with shifted pump resonances, we investigate the scaling of conversion efficiency. Integrating Eq. 1 we find the soliton energy as $E_{cs} = 4\sqrt{\Delta}$. Dividing the soliton energy with energy input per round trip, we find the soliton efficiency as

$$\eta_{cs} = \frac{\theta^2 E_{cs}}{\alpha^2 X\, T} = \frac{\theta^2 4\sqrt{\Delta}}{\alpha^2 X\, T}, \qquad (2)$$

which becomes $\eta_{cs} = \frac{16\sqrt{\Delta}}{X\, T}$ in absence of intrinsic losses, where $X$ is the normalized power, T is the normalized roundtrip time, $\theta$ is the coupling rate between bus and ring and $\alpha$ is the cavity loss per roundtrip. Note that $\eta_{cs}$ gives a good approximation of the conversion efficiency for our purposes, but it has a slightly different definition, where 100% soliton efficiency means that the pump has been transferred perfectly into a hyperbolic secant shape. Equation 2 should be treated as a rule-of-thumb approximation as it does not take into account pump saturation, thus higher than 100% conversion efficiency can be predicted, which is of course not realistic.

We plot the soliton efficiency with regards to detuning in Fig S1.d, with the waveforms of the closed form solutions for selected detuning values (labelled 1-3) plotted in Fig S1.e-f. The normalized roundtrip time of the cavity was the same as in Fig. 2b-c. The figures emphasize the dramatic improvement that can be made in conversion efficiency by decreasing the normalized power from 2.52 to 0.32 and by increasing the normalized detuning. Unfortunately, as was illustrated in Fig. 2b, such improvements are not available in standard anomalous-dispersion cavities (i.e. without shifted pump resonance). The reason for this limitation is that increased detuning leads to the pump being operated far away from resonance center such that it is inefficiently coupled into the cavity. This ultimately sets a maximum detuning (and minimum power) limit at $\Delta = \pi^2 X / 8$ as has been discussed in previous work[1,2]. The presence of modulational instability (MI) also eliminates the possibility to operate the DKS with the pump near the center of resonance. Thus, conversion efficiency is fundamentally limited in unperturbed anomalous-dispersion cavities, as was demonstrated by our simulations in Fig. 2b which found the peak conversion efficiency at $X = 2.52$ and $\Delta = 2.24$, with the simulated comb spectrum displayed in Fig S1a.

The purpose of shifting the pump resonance is to separate the detuning of the pump from the detuning of the comb. This way, we overcome the upper limit set on the comb detuning, allowing us to scale up the converted power by increasing the number of lines. A second consequence is that the DKS can be operated with the pump at the center of resonance at a significantly lower power, possibly due to reduced modulational instability (see section S2). The combination of these two factors allows the DKS to be operated with extremely high efficiency, as is predicted by Fig. S1d. This allows us to reach the low normalized power of $X = 0.32$ and a high detuning as is shown in Fig. S1b-c.

The soliton efficiency in equation 2 is also dependent on another parameter, the normalized roundtrip time defined as $T = \frac{1}{FSR} \sqrt{\frac{2\alpha}{|\beta_2|L}}$. Thus, it can be tuned in practice by changing the GVD ($\beta_2$). Tuning this parameter not only impacts the conversion efficiency, but also the number of lines. By doing Fourier analysis on Eq. 1 we find that the 3 dB drop off point of the soliton is found at by $f_{3db} = \frac{0.8814\sqrt{\Delta}}{\pi^2}$, where $f$ is the normalized frequency. The number of lines is found as $N = \frac{2f_{3db}}{f_{FSR}} + 1 = 2\frac{0.8814\sqrt{\Delta}}{\pi^2}T + 1$, where $f_{FSR} = 1/T$ is the normalized FSR. The added one line is to include the pump line which is in the middle of the spectrum.

It thus becomes clear that in an unperturbed anomalous-dispersion cavity, where the detuning is limited to $\Delta = \pi^2 X/8$, there will be a trade-off between the number of lines and conversion efficiency. The conversion efficiency [4] will scale as $\eta_{cs} \propto 1/N$ (since $N \propto T$ and $\eta_{cs} \propto 1/T$). However, this limitation is lifted when the pump resonance is allowed to be shifted, as we can select $\Delta$ more freely. With careful engineering, we can in principle find the correct combination of $\Delta$, $X$ and $T$ to generate high efficiency DKS for a pre-determined number lines as we find in section S3. It is then up to the designer to translate the normalized parameters into realistic designs depending on requirements and limitations of fabrication.

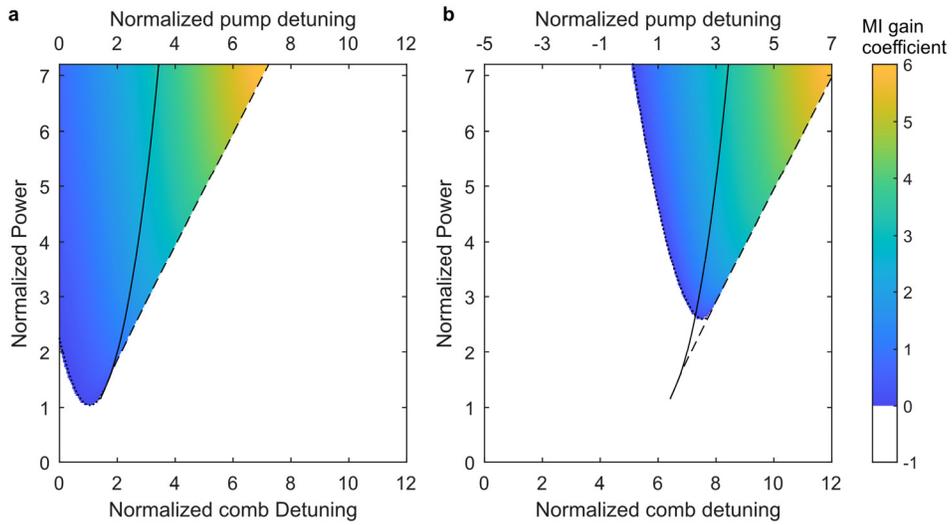

**Fig. S2 MI gain coefficient estimated from theory. a**, The color grading shows the maximum MI gain coefficient acquired from analytical equation for a cavity with an unshifted resonance. The dashed line and solid line confine the area of bistability. The dotted line and the dashed line confine the area where MI was found in simulations. **b**, Same as **a**, but with the pump resonances shifted by 5 normalized unit, same as in Fig.2c. The gain existence acquired from the analytical equation matches well the MI area found in our simulations.

## S2. Theoretical comparison of the MI existence area

A highly interesting result from Fig. 2c is that the modulational instability is limited to higher power levels. Here we analytically investigate these results using the gain coefficient derived for small electric field perturbations in a CW driven cavity[5]. The gain coefficient is written as

$$\lambda = -1 \pm [4(\Delta - \xi\omega^2)Y - (\Delta - \xi\omega^2)^2 - 3Y^2]^{1/2}, \qquad (2)$$

where $\xi$ is the sign of the GVD (set to -1 for anomalous dispersion), $\omega$ is the radial frequency and $Y$ is the normalized intracavity power. In the presence of a shifted pump resonance the detuning is set to equal the normalized comb detuning, $\Delta = \Delta_c$. We set the intracavity power by solving the CW steady-state cubic equation $X = Y^3 - 2\Delta_p Y^2 + (\Delta_p{}^2 + 1)Y$, where $\Delta_p$ is the normalized pump detuning. We use these equations to numerically find the maximum gain coefficient with regards to $\omega$ for the existence maps in Fig. 2b-c. Fig. S2a shows the results for the unshifted cavity from Fig. 2b, while Fig. S2b shows the shifted cavity from Fig. 2c, both showing excellent agreement with the simulations. These results show that the MI dynamics can be predicted using the pump detuning to define the CW intracavity build-up and the comb detuning to predict the interaction between comb lines. The high comb detuning results in reduced phase matching which limits the MI to higher normalized power levels. This absence of MI at lower power might be the reason why DKSs are allowed to exist near the center of pump resonance in Fig. 2c.

## S3. Scaling of high-conversion efficiency DKS microcombs

The existence map in Fig. 2c shows that the normalized power needs to be reduced significantly to reach high efficiency DKS-states. However, this does not restrict these waveforms to low input power, as the normalized power involves several parameters. In this section, we demonstrate via simulations how the conversion efficiency of the photonic molecule DKS microcombs can be maintained above 99% while increasing the pump power over 10 dB by simply adapting the comb detuning, pump detuning, GVD coefficient and coupling rate to the bus waveguide, where the latter two parameters can be engineered with lithographic control in silicon nitride. We also provide an experimental demonstration based on a different photonic molecule design that demonstrates that high conversion efficiency can be obtained with largely dissimilar power levels.

When increasing the input power, other parameters can be adjusted to maintain the same normalized values. This is demonstrated in Fig. S3a-b, which shows an example of how the comb can be operated at higher CW input power ($P_{in}$), while maintaining the same relative comb bandwidth and conversion efficiency, and drastically changing the pump power almost by two orders of magnitude. These combs were generated with the numerical model used in Fig.2, but with a higher shift to the pump resonance, allowing near 99% conversion efficiency. As the input power is increased (Fig.

S3a), the coupling, pump detuning, comb detuning and GVD are all increased in the same proportion to maintain the same normalized pump power.

Fig. S3c-d reveal that high efficiency solitons can be generated at different comb bandwidths. Fig. S3c shows the scaling of parameters as the input power is increased, with the GVD kept constant. The corresponding comb states demonstrate an increase in power and number of lines, maintaining the 99% conversion efficiency. Note that the parameter scaling does not keep the normalized parameters ($X$, $\Delta_c$, $\Delta_p$ and $t'$) fixed, instead a new combination is found in each case to allow the increased number of lines. This change is expected according to our analysis of number of lines in section S1.

The scaling in Fig. S3a and Fig. S3c shows that increasing the coupling factor can be modified to scale high efficiency comb states to a higher power level. To demonstrate this, we fabricated a second device with higher coupling rate by placing the main cavity closer to the bus waveguide, resulting in extrinsic Q-factor of 1-2 million. Using the same initiation methods as in Fig. 3, we generate a DKS state measured with 55% conversion efficiency (see Fig. S3e). The estimated on-chip CW power was 8 mW, resulting in roughly 2dB increase in total comb power compared to Fig. 3e.

The device used in the experimental demonstration of Fig.S3, had a facet coupling loss of ~2dB per facet. It had a main cavity of radius 227.26 µm and the radius of the auxiliary cavity was 23.36 µm. All waveguides had the same dimensions of 1800 nm width and 740 nm height. The gap between rings was 450nm, the gap between main cavity and bus was 300nm, and the gap between auxiliary cavity and bus was 670 nm. The main cavity was measured (referenced to 1565 nm) with GVD of $\beta_2 = -81\ ps^2/km$, FSR of 99.81 GHz, intrinsic Q factor with an average near 14 million, and a coupling rate to bus waveguide corresponding to extrinsic Q of 1-2 million near 1565nm. The maximal resonance shift induced by the auxiliary cavity was measured as 735 MHz. The auxiliary cavity resonance at 1565 which interacted with the main cavity had 0.6 million intrinsic Q and coupling rate to a bus waveguide corresponding to 50 million of extrinsic Q.

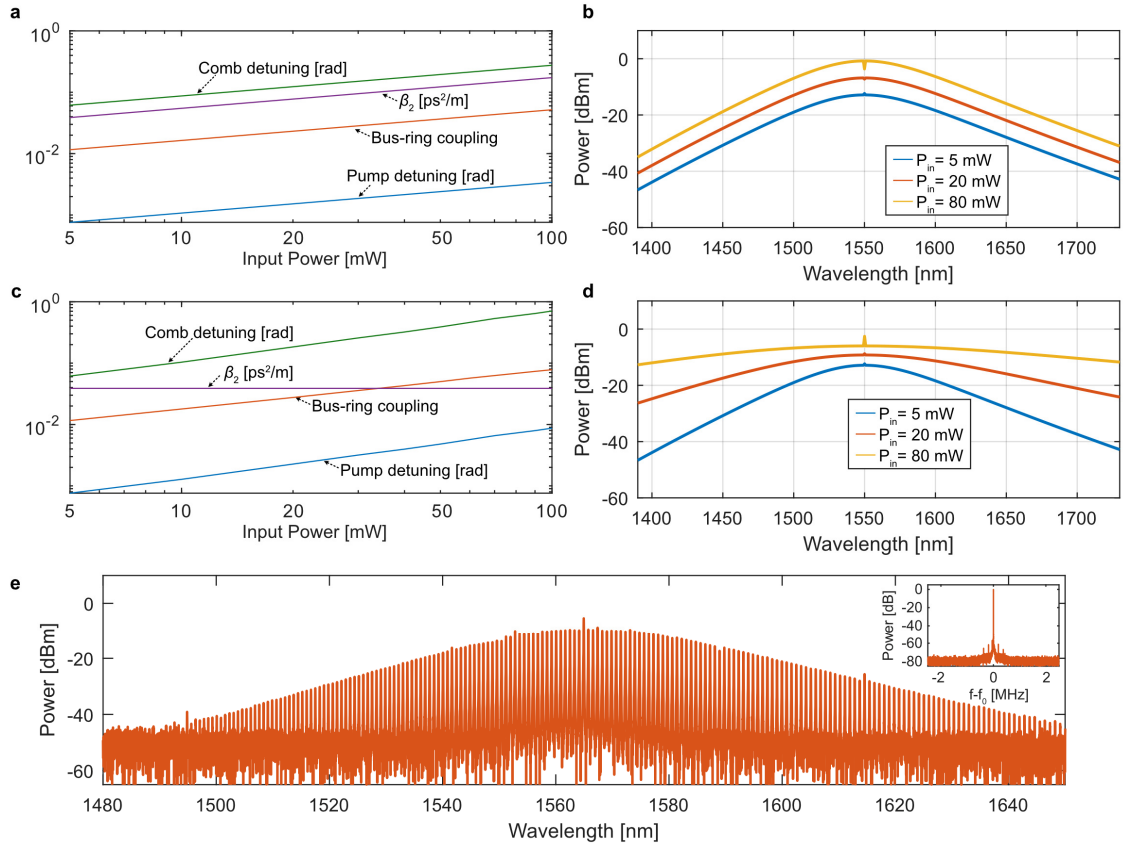

**Fig. S3 Scaling of high efficiency solitons. a**, For each input power level, the displayed parameters are scaled in order to maintain >99% conversion efficiency without changing the comb envelope, with three corresponding comb distribution shown in **b**. **c**, An example of parameter tuning to change comb bandwidth while maintaining 99% conversion efficiency, with three corresponding comb distributions shown in **d**. **e**, A comb spectrum acquired from a device which has slightly higher coupling factor between bus and ring compared to the device used in Fig. 3, resulting in 55% conversion efficiency. The inset shows the downconverted repetition-rate beatnote, measured with 100Hz radio bandwidth, indicating a coherent comb state.

## References

1. Coen, S. & Erkintalo, M. Universal scaling laws of Kerr frequency combs. *Opt. Lett.* **38**, 1790–1792 (2013).

2. Barashenkov, I. V & Smirnov, Y. S. Existence and stability chart for the ac-driven, damped nonlinear Schrödinger solitons. *Phys. Rev. E* **54**, 5707–5725 (1996).

3. Herr, T. *et al.* Temporal solitons in optical microresonators. *Nat. Photonics* **8**, 145–152 (2014).

4. Bao, C. *et al.* Nonlinear conversion efficiency in Kerr frequency comb generation. *Opt. Lett.* **39**, 6126–6129 (2014).

5.    Haelterman, M., Trillo, S. & Wabnitz, S. Dissipative modulation instability in a nonlinear dispersive ring cavity. *Opt. Commun.* **91**, 401–407 (1992).